\documentclass[12pt]{article}

\usepackage{graphics,epsfig}
\usepackage{amsmath}
\usepackage{amsfonts}
\usepackage{amssymb}
\newcommand{\be}{\begin{equation}}
\newcommand{\ee}{\end{equation}}
\newcommand{\ba}{\begin{eqnarray}}
\newcommand{\ea}{\end{eqnarray}}

\date{}

\begin{document}
\title{Chiral perturbation theory vs. Linear Sigma Model in a chiral imbalance medium}
\author{A. A. Andrianov$^{1,2}$
\footnote{E-mail: andrianov@icc.ub.edu},
\ V. A. Andrianov$^{1}$
\footnote{E-mail: v.andriano@rambler.ru},
\ D. Espriu$^{2}$
\footnote{E-mail: espriu@icc.ub.edu},
\\
\small{$^1$\ \  V. A. Fock Department of Theoretical Physics,}\\
\small{Saint-Petersburg State University, 199034, St. Petersburg, Russia}\\
\small{$^2$\ \ Departament  de F\'isica Qu\`antica i Astrof\`isica }\\ \small{and Institut de Ci\`encies del
Cosmos (ICCUB),}\\\small{
Universitat de Barcelona, Mart\'\i \ i Franqu\`es 1, 08028 Barcelona, Spain}}
\maketitle

%\keyword{ chiral imbalance; chiral perturbation theory; linear sigma model; charged pion decay in
%chiral medium; local parity breaking}

\abstract{
We compare the chiral perturbation theory (ChPT) and the linear sigma model (LSM) as realizations of low energy quantum chromodynamics (QCD)
for light mesons  in a  chirally-imbalanced medium.  The relations between the low-energy constants of the
chiral Lagrangian and the corresponding constants of the linear sigma model are established as well as
the expressions for the decay constant of $\pi $-meson in the medium and for the mass of the $a_0$. In the
large $N_c$ count taken from QCD the correspondence of ChPT and LSM is remarkably  good and provides a solid
ground for the search of chiral imbalance manifestations in pion physics. A possible experimental detection of chiral
imbalance (and therefore a phase with local parity breaking)  is outlined in the charged pion decays
inside the fireball.
}

\section{Introduction}
The possible generation of a phase with local parity breaking (LPB) in nuclear matter at extreme conditions
such as those reached in heavy ion collisions (HIC) at the Relativistic Heavy Ion Collider (RHIC) and the Large Hadron Collider (LHC) \cite{reviews} has been
examined recently \cite{kharzeev,anesp,avep}.
It has been suggested in \cite{kharzeev} that at increasing temperatures an isosinglet pseudoscalar background could
arise due to large-scale topological charge fluctuations (studied recently in lattice quantum chromodynamics (QCD) simulations
\cite{lattice}).

These considerations led eventually to the observation of the so-called chiral magnetic effect
(CME)~\cite{kharzeev}
in the STAR and PHENIX experiments at RHIC \cite{star}. The effect should be most visible
for non-central HIC where large angular momenta induce large magnetic fields contributing
to the chiral charge separation. {However, the CME  may be only a partial explanation of the STAR and PHENIX experiments and other backgrounds play a comparable role (see the reviews \cite{chiraleff}).  In a recent report~\cite{measuCME} the measurements of the chiral magnetic effect in Pb–Pb collisions with A Large Ion Collider Experiment (ALICE) were estimated and perspectives to improve their precision in future LHC runs were outlined}.

 For central collisions it was proposed in \cite{anesp} that
the presence of a phase where parity was spontaneously broken could be a rather generic feature of QCD.  Local parity breaking can be induced by difference between the densities of the right- and
left-handed chiral fermion fields (chiral Imbalance) {in metastable domains with non-zero topological charges. Thus our analysis concerns solely
the events in the central heavy ion collisions where the magnetic fields are negligible. It is seen in the experiments~\cite{star,chiraleff}, and was also found in lattice QCD (see~\cite{lattice}). The validity of CME and its percentage in observations is
well analyzed in \cite{measuCME}.
Thereby the elimination of electromagnetic effects %previously e.m., confirm original meaning is retained
 is justified and allows to measure solely the chiral chemical potential without contamination by magnetic fields and related~backgrounds.}

In the hadron phase we shall assume that as a consequence of topological charge fluctuations, the environment
in the central HIC generates a pseudoscalar background growing approximately linearly in time.
This background is associated with a constant axial vector whose zero component is identified with a chiral
chemical potential. In such an environment one could search for a possible manifestation of LPB {in dilepton
probes.}   In particular, in \cite{aaep} it was shown that a good part of the excess of dileptons produced in
central heavy-ion collisions \cite{phenix} might be a consequence of LPB due to the generation
of a pseudoscalar isosinglet condensate whose precise magnitude and time variation depends on
the dynamics of the HIC.

The complete description of a medium with chiral imbalance should also take into account thermal fluctuations of the medium. In this paper the description in a zero temperature limit is considered and to understand the changes for non-zero temperatures we rely on the results of lattice computations of quark matter with chiral imbalance and a temperature of order 150 MeV undertaken \cite{braguta}. Thus our calculations keep the tendency of increasing chiral condensate and decreasing pion masses when the temperature grows.

This paper is mostly concerned with the possibility of identifying LPB in the hadron phase of QCD in HIC. Such a
medium would be simulated by a chiral chemical potential $\mu_5$. Adding to the QCD Lagrangian the term
$\Delta{\mathcal
L}_q=\mu_5 q^\dagger \gamma_5 q \equiv \mu_5 \rho_5$, we allow for non-trivial
topological fluctuations \cite{aaep} in the nuclear (quark) fireball, which are ultimately related to fluctuations
of gluon fields. The transition of the quark--gluon medium characteristics to a hadron matter reckons on the
quark--hadron continuity~\cite{contin} after hadronization of quark--gluon plasma. The behavior of various spectral
characteristics for light scalar and pseudoscalar ($\sigma, \pi^a, a^a_0$)-mesons by means of a QCD-motivated
$\sigma$-model Lagrangian was recently derived  for $SU_L (2) \times SU_R (2)$ flavor symmetry including an
isosinglet
chiral chemical potential \cite{eqcd16}. The structural constants of the $\sigma$-model Lagrangian were taken as
input
parameters suitable to describe the light meson properties in vacuum and then they are extrapolated to a chiral medium.
In this way ad hoc there is no reliable predictability in the determination of the hadron system response on
chiral imbalance, and reaching quantitative predictions requires a phenomenologically justified hadron dynamics.
To increase predictability, we  extend the vacuum chiral Lagrangians \cite{GL,kais} with phenomenological
low-energy
structural constants taking into account the chiral medium in the fireball with a chiral imbalance.
It is shown that $\sigma$-model parametrization of \cite{eqcd16} fits well the pion phenomenology at low energies
as derived from ChPT.

Next it is  described how pions modify their dynamics in decays in a chiral medium, in particular, charged pions
stop
decaying into muons and neutrinos for a large enough chiral chemical potential. A possible experimental detection of chiral
imbalance (and therefore a phase with local parity breaking)  is outlined in the charged pion decays
inside the fireball.
\section{Chiral Lagrangian with Chiral Chemical Potential}
The chiral Lagrangian for pions  describing their mass spectra and decays in the fireball with a chiral imbalance
can
be implemented with the help of softly broken chiral symmetry in QCD transmitted to hadron media, a properly constructed  covariant derivative:
\begin{equation}
D_\nu \Longrightarrow  \bar D_\nu - i \{{\bf I}_q\mu_5 \delta_{0\nu}, \star
\}={\bf I}_q \partial_\nu - 2i{\bf I}_q\mu_5 \delta_{0\nu},\label{covderu}
\end{equation}
where we skipped the electromagnetic field. The axial chemical potential is introduced as a constant time component
of an
isosinglet axial-vector field.

In the framework of large number of colors $N_c$ \cite{kais} the SU(3) chiral Lagrangian in the strong interaction
sector
contains the following dim=2 operators \cite{kais},
\be
{\cal L}_2=\frac{F^2_0}{4} <- j_\mu j^\mu  +\chi^\dagger U+ U^\dagger \chi >,
\label{L2lag}
\ee
where $<...>$ denotes the trace in flavor space, $j_\mu \equiv U^\dagger \partial_\mu U $, the chiral
field $U=\exp(i \hat\pi/ F_0)$,\ the bare pion decay constant $F_0 \simeq 92$ MeV, $\chi(x)=2B_0 s(x)$ and
$M_\pi^2=2B_0 \hat m_{u,d}$,  the tree-level neutral
pion mass. The constant $B_0$ is related to the chiral quark condensate $<\bar q q>$ as $ F_0^2 B_0 = - <\bar q
q>$.
Taking now the covariant derivative in \eqref{covderu} it yields
\begin{equation}
{\cal L}_2 (\mu_5)={\cal L}_2(\mu_5=0)+\mu_5^2 N_f F_0^2.
\end{equation}

Herein we have used the identity for $U\in SU(n)$,$< j_\mu>=0$.
In the large $N_c$ approach the dim=4 operators \cite{kais} in the chiral Lagrangian are given by
\ba
{\cal L}_4 =\bar L_3  <j_\mu j^\mu j_\nu j^\nu> + L_0 <j_\mu j_\nu j^\mu j^\nu> - L_5 < j_\mu j^\mu (\chi^\dagger
U+ U^\dagger \chi)>,\label{L4lag}
\ea
where $L_0, \bar L_3, L_5$ are bare low energy constants.
For SU(3) and SU(2) $<j_\mu>=0$ and there is the identity
\ba <j_\mu j_\nu j^\mu j^\nu>=-2 <j_\mu j^\mu j_\nu j^\nu> + \frac12 <j_\mu j^\mu> <j_\nu j^\nu> +  <j_\mu j_\nu>
<j^\mu j^\nu> , \label{id1}\ea
whereas for SU(2) there is one more identity
\be 2 <j_\mu j^\mu j_\nu j^\nu> = <j_\mu j^\mu> <j_\nu j^\nu> . \label{id2}\ee

Applying these identities one finds the four-derivative Gasser--Leutwyler (GL) operators for the SU(3) chiral
Lagrangian
\ba{\cal L}_4 &=& L_1 <j_\mu j^\mu> <j_\nu j^\nu> +  L_2 <j_\mu j_\nu> <j^\mu j^\nu> +L_3  <j_\mu j^\mu j_\nu
j^\nu>\nonumber\\&&
- L_5 < j_\mu j^\mu (\chi^\dagger U+\chi U^\dagger)>\ea with
\be L_1=  \frac12 L_0;\ L_2= L_0;\ L_3 =\bar L_3 -2L_0.\ee

For SU(2) one has a further reduction of the dim=4 Lagrangian,
\be {\cal L}_4 = \frac14 l_1 <j_\mu j^\mu> <j_\nu j^\nu> + \frac14 l_2 <j_\mu j_\nu> <j^\mu j^\nu>- \frac14 l_4 <
j_\mu j^\mu (\chi^\dagger U+ U^\dagger \chi)> \label{final}\ee
where $l_{1}, l_{2}, l_{4}$ are renormalized SU(2) low energy constants (as compared to \cite{GL} )
with normalization so that
\be  l_1 = 2 L_0 +2 \bar L_3,\  l_2 =  4 L_2= 4L_0,\  (l_1 +l_2) = 2\bar L_3 + 6 L_0;\quad l_4 = 4 L_5.\ee
We stress that this chain of reductions is valid only if $<j_\mu>=0$.

The response of the chiral Lagrangian on chiral imbalance is derived with the help of the covariant derivative
\eqref{covderu}
applied to the Lagrangian \eqref{L4lag},
\ba
&&\Delta {\cal L}_4 (\mu_5) =
-\mu_5^2\{ 12(l_1 + l_2) <j^0 j^0> -4(l_1+l_2) <j_k j_k> - l_4 <\chi^\dagger U+ U^\dagger  \chi >\}.
\ea
We notice that this result is drastically different from what one could obtain from the final Lagrangian~\eqref{final}.
This is because the identities \eqref{id1} and \eqref{id2} are violated if $<j_\mu>\not=0$.
The above modifications change differently the coefficients in the dispersion law in energy $p^0$ and
three-momentum $|\vec p|$ for the mass shell  as well as modify the mass term for pions (all together it gives the
inverse propagator of pions),
\be
{\cal D}^{-1}( \mu_5) = (F_0^2 + 48 \mu^2_5 (l_1 + l_2)) p_0^2 -(F_0^2 +16 \mu^2_5 (l_1+l_2) )|\vec p|^2 - (F^2_0
+ 4l_4\mu_5^2) m_\pi^2(0) \to 0. \label{massshell}
\ee

In the leading order of large $N_c$ expansion  the empirical values of
the SU(2)
Gasser-Leutwyler (GL) %please define if necessary.
 constants $l_1, l_2, l_4$ are given in \cite{GL}.
%\ba
%&&l_1^r = (-0.4 \pm 0.6)\times 10^{-3};\ l_2^r = (8.6 \pm 0.2) \times 10^{-3};\nonumber\\&& l_1^r + l_2^r = (8.2
%\pm 0.8) \times 10^{-3};\
%l_4^r =  (2.64 \pm 0.01)\times 10^{-2}.
%\ea

%They can be obtained also if they are normalized at the renormalization group scale $\mu \simeq M_\pi \simeq 140$~MeV,
%$\log\Big(m_\pi/\mu\Big)\simeq 0$ .

Thus in the pion rest frame
\be
F_\pi^2(\mu^2_5)\simeq F_0^2 + 48 \mu^2_5 (l_1 + l_2);\quad m_\pi^2(\mu^2_5)\simeq \Big(1 -4 \frac{\mu_5^2}{F_0^2}
(12(l_1 + l_2) -  l_4  )\Big) m_\pi^2 (0),\label{fandm}
\ee
i.e., the pion decay constant is growing and its mass is decreasing in the chiral media. From these relations one can determine the quark condensate dependence on the chiral imbalance,
\be
<\bar q q>(\mu_5)  = <\bar q q>(0)\Big(1 + 4 l_4 \frac{\mu_5^2}{F^2_0
}\Big),
\ee
wherefrom one concludes that for $l_4 > 0$ \cite{GL} the magnitude of quark condensate increases with growing the chiral chemical potential as it was predicted in \cite{braguta, eqcd16, eqcd2017}

\section{Linear Sigma Model for Light Pions and Scalar Mesons in the Presence of Chiral Imbalance: Comparison to
ChPT}
Let us compare these constants with those ones estimated from the linear sigma model (LSM) built in~\cite{eqcd2017,efflag}.
The sigma model was build with realization of SU(2) chiral symmetry to describe pions and isosinglet and
isotriplet scalar mesons.
Its Lagrangian reads
\begin{eqnarray}%\mathbf
{L}&=& N_c \Big\{\frac{1}{4}\,<(D_{\mu}H\,(D^{\mu}H)^{\dagger}>
+\frac{B_0}{2}\,  <m(H\,+\,H^{\dagger}>
+\frac{M^{2}}{2}\,<HH^{\dagger}>
\nonumber \\
\,&-&\frac{\lambda_{1}}{2}\,<(HH^{\dagger})^{2}>
-\frac{\lambda_{2}}{4}\, <(HH^{\dagger})>^{2}
+\frac{c}{2}\,(\det H +\,\det H^{\dagger})\Big\},
\label{lagr_sigma}
\end{eqnarray}
where $H = \xi\,\Sigma\,\xi$ is an operator for meson fields, $N_c$ is a number of colours,
\(m\) is an average mass of current $u,d$ quarks,
\(M\) is a `'tachyonic'' mass generating the spontaneous breaking of chiral symmetry,
\(B_0, c,\lambda_{1},\lambda_{2}\) are real constants.

The matrix \(\Sigma\) includes the singlet scalar meson \(\sigma\), its vacuum average \(v\) and the isotriplet of
scalar mesons \( a^{0}_{0},a^{-}_{0},a^{+}_{0}\), the details see in ~\cite{eqcd2017,efflag}. The covariant
derivative
of $H$ including the chiral chemical potential $\mu_5$ is defined in \eqref{covderu}.
The operator \(\xi\) realizes a nonlinear representation (see \eqref{L2lag}) of the chiral group $SU(2)_L
\times SU(2)_R$,
namely,  $\xi^2 = U$.

The diagonal masses for scalar and pseudoscalar mesons read
\begin{equation}
\begin{aligned}
m^{2}_{\sigma} &=-2\left( M^{2}-6\,(\lambda_{1}+\lambda_{2})F_\pi^{2}+c+2\mu^2_5 \,\right)\\
m^{2}_a &=-2 \left( M^{2}-2\,(3\lambda_{1}+\lambda_{2})F_\pi^{2}-c+2\mu^2_5 \, \right)\\
m^{2}_{\pi} (\mu_5) &=\frac{2\,b\,m}{F_\pi} \simeq m^{2}_{\pi} (0)\left(1 -
\dfrac{\mu^{2}_{5}}{2(\lambda_{1}+\lambda_{2})F_0^2}\right) \\
F^2_\pi(\mu_{5}) &=\dfrac{M^{\,2}+2\mu^{2}_{5}+c}{2(\lambda_{1}+\lambda_{2})} = F_0^2 +
\dfrac{\mu^{2}_{5}}{\lambda_{1}+\lambda_{2}}.
\end{aligned}
\end{equation}

From spectral characteristics of scalar mesons in vacuum one fixes the Lagrangian parameters,
\(\lambda_{1}=16.4850\), \(\lambda_{2}=-13.1313\), \(c=-4.46874\times10^4~{\text{MeV}}^2\), \( F_0 = 92 \text{MeV}\), \(b = B_0 F_0  =1.61594\times10^5~{\text{MeV}}^2\) \cite{eqcd2017}.

The change of the pion-coupling constant $F_0$ is determined by potential parameters as compared to the ChPT
definition,
\be
{\Delta F_\pi^2\over\mu_5^2} = \frac{1}{\lambda_1 +\lambda_2}\approx 0.3\quad \mbox{\it vs}\quad 48  (l_1 + l_2),
\ee
i.e. $(l_1 + l_2)\approx 6.2\times 10^{-3}$. It is a satisfactory correspondence to the pion phenomenology \cite{GL} .

Analogously, in the rest frame using the pion mass correction,\\ $ m^2_\pi(\mu_5)F^2_\pi(\mu_5) \simeq 2m_q b
F_\pi(\mu_5) $ it
is easy to find the estimation for
 \be l_4 = \frac{1}{8(\lambda_1 +\lambda_2)})\approx 3.7\times
 10^{-2}, \ee
 wherefrom one can also guess the relation $6(l_1 + l_2) = l_4$ following from the LSM. It is  again a satisfactory correspondence to the pion phenomenology \cite{GL} .

For moving mesons with $|\vec p|\not = 0$ and the CP %please define if necessary
 breaking mixing of scalar and pseudoscalar mesons the
effective masses $m^{2}_{eff\mp}$ take the form,
\begin{gather}
m^{2}_{eff-}=\frac{1}{2}\left( 16 \, \mu^{2}_{5} +m^{2}_a +m^2_\pi - \sqrt{(16 \mu^{2}_{5}+m^{2}_{a}+m^2_\pi)^{2}
-4 \left(  m^{2}_a \,m^2_\pi -16 \mu^{2}_{5}  \,|\vec p|^{2}  \right) } \,\ \right),
\nonumber \\
m^{2}_{eff+}=\frac{1}{2}\left( 16 \, \mu^{2}_{5} +m^{2}_a +m^2_\pi + \sqrt{(16 \mu^{2}_{5}+m^{2}_{a}+m^2_\pi)^{2}
-4 \left(  m^{2}_a \,m^2_\pi -16 \mu^{2}_{5}  \,|\vec p|^{2}  \right) } \,\ \right).
\end{gather}

For small $\mu^{2}_{5}, m^2_\pi \ll m^{2}_a \simeq 1 GeV^2$ one can approximate the dependence on the wave vector
$\vec p$
\be
m^{2}_{eff-} \simeq m^2_{\pi} - 16 \mu_5^2 \frac{|\vec p|^2}{m_a^2}.
\ee

Comparing with \eqref{massshell}
one establishes the relationship of isotriplet scalar mass and GL constants
\be
m_a = \frac{F_0}{\sqrt{2(l_1 + l_2)}} \simeq 0.9 GeV,    \label{ma}
\ee
which is close to the Particle Data Group %please define if necessary
 value within the experimental error bars~\cite{PDG}.
\section{Possible Experimental Detection of Chiral
Imbalance in the Charged Pion Decays}
The predicted distortion of the mass shell condition can be detected in decays of charged pions  when the effective
pion mass approaches the muon mass.
Let us find the threshold value for the $\pi^+\to \mu^+\nu$ decay.
If a charged pion was generated in chiral medium its mass is lower than in the vacuum and the condition for its
decay follows from \eqref{massshell},
\be
\left(1-32 (l_1 + l_2)  \frac{\mu_5^2}{F_0^2}\right) |\vec p|^2 + \left(1-24 (l_1 + l_2)  \frac{\mu_5^2}{F_0^2}\right)m^2_{0,\pi}  \geq |\vec p|^2
+m^2_\mu ,
%\quad
%\frac{m^2_a}{16\mu_5^2} \geq \frac{|\vec p|^2 + m^2_{0,\pi}}{m^2_{0,\pi} - m^2_\mu },
\ee
where we have used the relations  $6(l_1 +l_2) \simeq l_4$. The decay channel is closed
for $|\vec p|^2 \simeq 0$  if $\mu_5 \simeq 160$ MeV. It must be detected as a substantial decrease of muon flow
originated from
pion decays in the fireball. When considering the decay process of a charged pion into a muon $+$ neutrino at values of the chiral chemical potential lower than $\mu_5 \simeq 160$ MeV  then still the muon yield from the fireball obviously decreases at sufficiently large momenta. It gives one a chance to measure the magnitude of
chiral chemical potential for sufficiently high statistics.

\section{Results}
\begin{itemize}
\item  For light mesons in the chiral imbalance medium we compared the chiral perturbation theory (ChPT) and the
linear sigma model (LSM)
as realizations of low energy QCD.
The relations between the low-energy constants of the chiral Lagrangian and
the corresponding constants of the linear sigma model are established and
expressions for the decay constant of the pion in the medium and the mass of the $a_0$
meson are found.
\item The low energy QCD correspondence of ChPT  and LSM
in the large $N_c$ limit is satisfactory and provide a solid ground for the search of chiral imbalance manifestation
in pion physics at HIC.
\item The resulting  dispersion law for pions in the medium allows us
reveal the threshold of decay of a charged pion into a muon and neutrino which can be
suppressed by increasing chiral chemical potential.
\item  As it is shown in \cite{eqcd2017}, at higher energies exotic decays of isoscalar mesons into three pions arise  due to
mixing of $\pi$ and $a_0$ meson states in the presence of chiral imbalance. It was  shown \cite{aaep,eqcd16,eqcd2017} that for a wider class of direct parity breaking at higher energies, in the framework of linear sigma model with isotriplet scalar ($a_0$) and pseudoscalar (pions) mesons, their mixing arises with the generation of $\pi\pi$ and $\pi\pi\pi$ decays of a heavier scalar state. Also, the independent check of our estimates could be done by lattice computation (cf.\cite{braguta}).
\item  A manifestation for LPB can also happen in the presence of chiral imbalance in the sector
of $\rho$ and $\omega$ vector mesons \cite{anesp,avep} and in this case the Chern--Simons
interaction plays a major role. It turns out \cite{anesp} that the spectrum
of massive vector mesons splits into three components with different polarizations $\pm,long$ having different
effective
masses $m_{V,+} < m_{V,long} < m_{V, -}$.
\item Thus a possible experimental detection of chiral imbalance in medium
(and therefore a phase with LPB) in the charged pion
decays and vector meson polarizations inside of the fireball can be~realized.
\item We would like to mention the recent proposal to measure the photon
polarization asymmetry in $\pi\gamma$ scattering \cite{harada,eqcd2017,ppn2019} as a way to detect LPB due to
chiral imbalance. This happens in the ChPT including electromagnetic fields
due to the Wess--Zumino--Witten operators.
\item One may be concerned about the appearance of changes in the properties of muons and neutrinos in the medium, but in our opinion,  this does not change the main estimates in Eq. (22), as a possible influence of chiral chemical potential
on lepton properties would be controlled by extra power of the Fermi coupling constant, i.e., by the next order in weak interactions with little hope to register it.
\item  We emphasize the similarity of our model results to lattice computation in \cite{braguta}: to the same tendency of increasing chiral
condensate and decreasing of pion mass when $ \mu_5$ grows for fixed temperatures about 150 MeV. It gives us the confidence (see \cite{eqcd16, eqcd2017})
that our spectral predictions are robust in a range of temperatures. We understand that for a more realistic quantitative description of the phenomena under discussion, thermal  effects, smearing of data and detector acceptance must be taken into account, which will be done in subsequent works with an extended team including the experimentalists.
\item Last  decade, different controversial conclusions on thermodynamics of quark matter with a chiral imbalance appeared  based on
different models of the Nambu-Jona-Lasinio (NJL) %please define if necessary.
 type. Among them, an opposite decreasing behavior of quark condensate in \cite{wrongcutoff}
was found due to an erroneous use of UV regularization in a NJL type model which mimicked chiral symmetry breaking, and chiral imbalance with chiral chemical potential being included into an UV cutoff. This kind of mistake in applications of NJL models has been known since the 1980s. The correct regularization based on vacuum definitions of cutoffs is elucidated in \cite{correctcutoff}.
In the treatment in  \cite{braguta}  based on lattice computations as well as in meson Lagrangians \cite{efflag}  where the UV finite chirally-symmetric computations are used, the problem is thoroughly resolved.
\end{itemize}
\section*{Acknowledgments}
This research was funded by the Grant FPA2016-76005-C2-1-P and Grant
2017SGR0929 (Generalitat de Catalunya),
  by Grant RFBR 18-02-00264 and by  SPbSU travel grants Id: 43179834, 43178702, 43188447, 41327333, 36273714, 41418623.

We express our gratitude to Angel G\'omez Nicola for stimulating discussions of how to implement chiral imbalance in ChPT.

%\conflictsofinterest{The authors declare no conflict of interest.}

%\reftitle{References}

\end{document}